\documentclass[letterpaper,10pt]{article} 

\usepackage{osameet3} 

\usepackage{amsmath,amssymb}
\usepackage[colorlinks=true,bookmarks=false,citecolor=blue,urlcolor=blue]{hyperref} 
\usepackage{wrapfig}
\usepackage{caption}
\usepackage{multicol}

\begin{document}

\title{Record Capacity-Reach of C band IM/DD Optical Systems over Dispersion-Uncompensated Links}

\vspace{-0.3cm}
\author{Haide Wang$^1$, Ji Zhou$^{1, \ast}$, Jinlong Wei$^2$, Wenxuan Mo$^1$, Yuanhua Feng$^1$, Weiping Liu$^1$, Changyuan Yu$^3$, and Zhaohui Li$^{4, 5}$}
\address{$^1$Department of Electronic Engineering, Jinan University, Guangzhou, China\\
$^2$Huawei Technologies Duesseldorf GmbH, European Research Center, Germany\\
$^3$Department of Electronic and Information Engineering, Hong Kong Polytechnic University, Hongkong, China\\
$^4$Guangdong Provincial Key Labratory of Optoelectronic Information Processing Chips and Systems, Sun Yat-sen University, Guangzhou 510275, China\\
$^5$Southern Marine Science and Engineering Guangdong Laboratory (Zhuhai), 519000, China}
\vspace{-0.1cm}
\email{E-mail: zhouji@jnu.edu.cn}

\copyrightyear{2021}

\vspace{-0.1cm}
\begin{abstract}
We experimentally demonstrate a C band 100Gbit/s intensity modulation and direct detection entropy-loaded multi-rate Nyquist-subcarrier modulation signal over 100km dispersion-uncompensated link. A record capacity-reach of 10Tbit/s$\times$km is achieved.
\end{abstract}

\vspace{0.1cm}
\section{Introduction}
\vspace{-0.05cm}
Due to the advantages of low cost, small footprint and low power consumption, intensity modulation and direct detection (IM/DD) systems are more suitable for short-reach transmissions than coherent systems. However, the achievable reach of IM/DD optical systems is limited by chromatic dispersion (CD). There is a lot of literature on the digital signal processing (DSP) to solve the CD-caused distortions in IM/DD optical systems.
\begin{wrapfigure}{r}{6cm}
	\vspace{-0.5cm}
	\includegraphics[width=\linewidth]{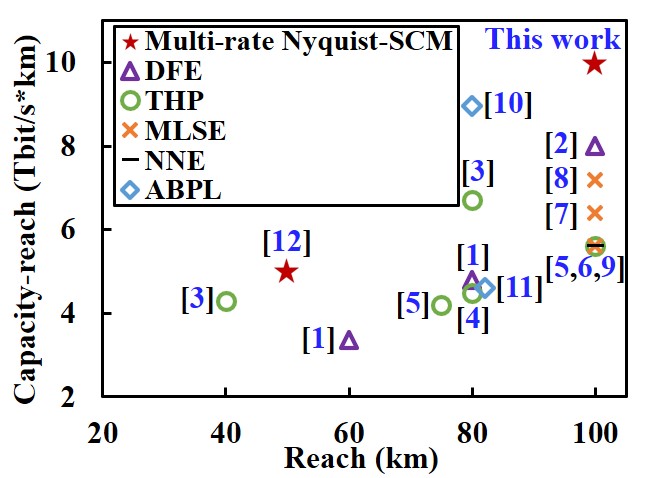}
	\caption{Capacity–reach of C band IM/DD optical systems.}
	\label{review}
\end{wrapfigure}
Fig. \ref{review} summarizes the capacity-reach for C band IM/DD optical systems over dispersion-uncompensated links. The most commonly used DSP algorithms include the decision feedback equalizer \cite{DFE1, DFE2}, Tomlinson–Harashima precoding \cite{THP1, THP2, THP3}, maximum likelihood sequence estimation (MLSE) \cite{MLSE1, MLSE2, MLSE3}, neural network equalizer (NNE) \cite{NNE} for the single-carrier modulation and adaptive bit and power loading (ABPL) \cite{APBL1, APBL2} for the multi-carrier modulation. In our previous work, we proposed a 100Gbit/s multi-rate Nyquist-subcarrier modulation (SCM) signal with seven subcarriers over 50km standard single-mode fiber (SSMF) transmission \cite{SCM}. In this paper, we extend the work and demonstrate a C band IM/DD 100Gbit/s clipped multi-rate Nyquist-SCM signal with sixteen subcarriers over 100km SSMF transmission. A record capacity-reach of 10Tbit/s$\times$km is achieved.

\section{DSP and Experimental Setups}
\vspace{-0.05cm}
\begin{figure}[!b]
	\vspace{-0.5cm}
	\centering
	\includegraphics[width=\linewidth]{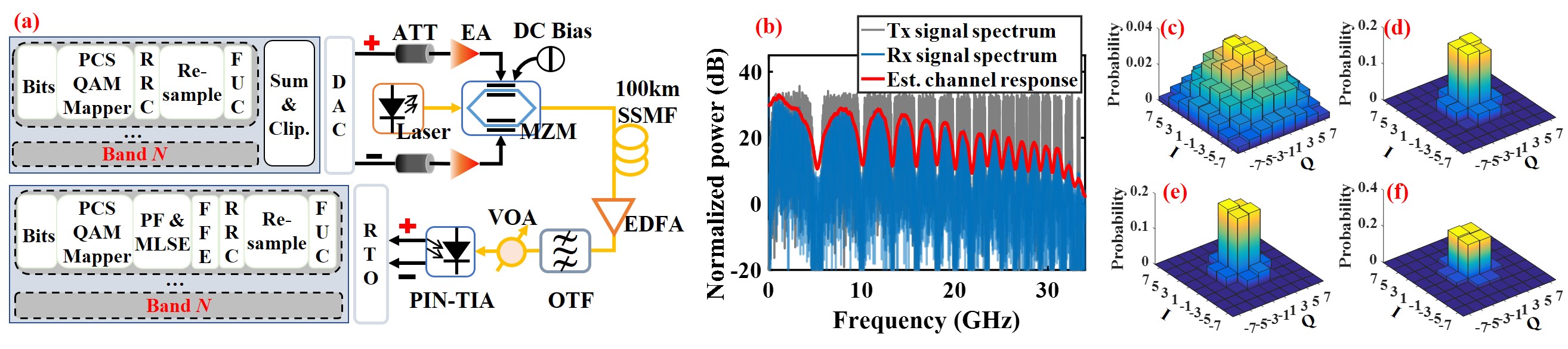}
	\caption{(a) Experimental setups of the C band IM/DD multi-rate Nyquist-SCM siganl over 100km dispersion-uncompensated link. (b) Spectrums of the transmitted and the received signal after 100km SSMF transmission. (c-f) PDFs of the recovered PCS 64QAM symbols on the 1$^{\rm st}$, 5$^{\rm th}$, 8$^{\rm th}$ and 12$^{\rm th}$ band, respectively.}
	\label{DSP}
\end{figure}
The experimental setups of the multi-rate Nyquist-SCM siganl over 100km dispersion-uncompensated link is shown in Fig. \ref{DSP}(a). The transmitter DSP of each band mainly includes the probabilistic constellation shaping (PCS) QAM mapping, rooted raised cosine (RRC) shaping, resampling and frequency up-conversion (FUC). Then the multi-rate Nyquist-SCM siganl is generated after the $N$-band signals superposition and clipping, where $N = 16$ in the experiment. PCS 64QAM is employed on the first thirteen bands and binary phase shift keying is used on the last three bands. Then the signal is up-loaded into a 90GSa/s digital-to-analog converter (DAC) with 3dB-bandwidth of 16GHz. The differential output of the DAC is amplified by electrical amplifiers (EA) with 6dB attenuator (ATT) and modulated on an optical carrier at 1550.02nm by a 40Gbps Mach-Zehnder modulator (MZM) @ push-pull mode. A 2.45V direct current (DC) bias is applied on the MZM. Then the optical signal is launched into the 100km SSMF with about 6.89dBm launch power. After 100km SSMF transmission, the optical signal is amplified to 9.9dBm by an erbium-doped fiber amplifier (EDFA) and filtered by an optical tunable filter (OTF). The received optical power (ROP) is adjusted by a variable optical attenuator (VOA). Then the optical signal is converted to eletrical signal by a 31GHz PIN-TIA with differential output and digitalized by a 80GSa/s real-time oscilloscope (RTO) with 36GHz cut-off bandwidth. Finally, the signal on each band is recovered by the receiver DSP, including frequency down-conversion (FDC), resampling, RRC filter, feedforward equalizer (FFE), two-tap post filter (PF), MLSE and QAM de-mapping.

\begin{figure}[!t]
	\centering
	\includegraphics[width=\linewidth]{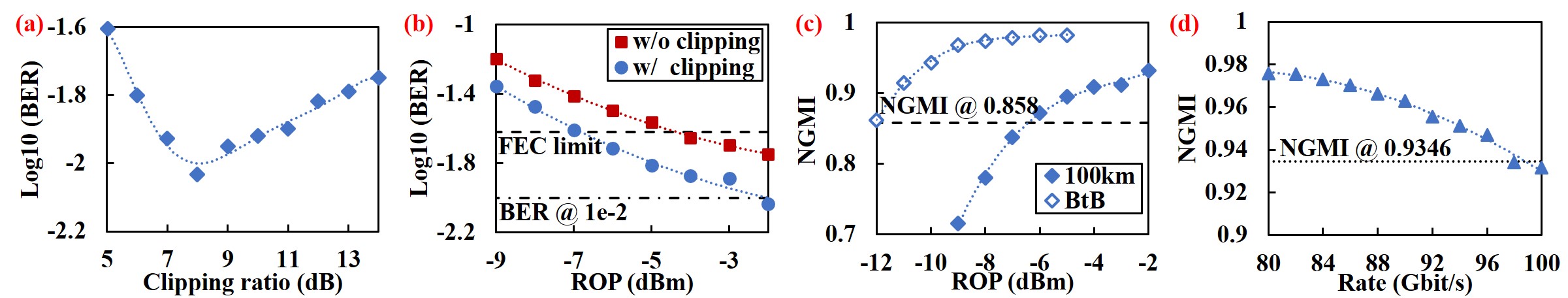}
	\caption{(a) BER of 100Gbit/s signal over 100km SSMF transmission versus clipping ratio. (b) BER of the 100Gbit/s signal over 100km SSMF transmission. (c) NGMIs of the 100Gbit/s signal over BtB and 100km SSMF transmission. (d) NGMI of the entropy-loaded and clipped signal at different data rates over 100km SSMF transmission.}
	\vspace{-0.28cm}
	\label{result}
\end{figure}
\section{Results and Discussions}
\vspace{-0.05cm}

The spectrums of the transmitted and the received signal over 100km SSMF transmission are shown in Fig. \ref{DSP}(b). The red curve is the estimated channel response. The ROP is $-2$dBm. The signal after 100km SSMF transmission suffers from not only the limited bandwidth but also the CD-caused power fading. Fortunately, the multi-rate Nyquist-SCM signal keeps away from the CD-caused 15 spectral nulls. As a result, the signal on each band can be recovered by a FFE, a two-tap PF and MLSE with one memory length. Figs. \ref{DSP}(c-f) show the probability distribution functions (PDFs) of the recovered PCS 64QAM symbols on the 1$^{\rm st}$, 5$^{\rm th}$, 8$^{\rm th}$ and 12$^{\rm th}$ band, respectively. The probability of the outer constellation points is small, while that of the inner is large. 

Bit error rate (BER) of 100Gbit/s signal over 100km SSMF transmission versus clipping ratio at ROP of $-2$dBm is shown in Fig. \ref{result}(a). The unclipped signal has a 14.03dB peak-to-average power ratio. BER is below $10^{-2}$ when the signal is clipped with optimal clipping ratio of 8dB. BER of the 100Gbit/s Nyquist-SCM signal without or with clipping over 100km SSMF transmission is shown in Fig. \ref{result}(b). BER of the signal without and with clipping is below the 20\% soft-decision forward error correction (SD-FEC) limit at ROP of about $-4.2$dBm and $-6.8$dBm, respectively. Normalized generalized mutual information (NGMI) thresholds of 7\% hard-decision FEC (HD-FEC) and 20\% SD-FEC are considered to be 0.9346 and 0.858, respectively. As Fig. \ref{result}(c) shows, the average NGMIs of the 100Gbit/s Nyquist-SCM signal over back-to-back (BtB) and 100km SSMF transmission are larger than 0.858 when ROP is larger than $-12$dBm and $-6$dBm, respectively. NGMI is 0.932 at ROP of $-2$dBm. Fig. \ref{result}(d) shows NGMI of the entropy-loaded signal at different data rates over 100km SSMF transmission. With the increase of data rate, NGMI decreases. NGMI is larger than 0.9346 when data rate is less than about 98Gbit/s. 

\section{Conclusion}
\vspace{-0.05cm}
By keeping away from the CD-caused spectral nulls, the entropy-loaded multi-rate Nyquist-SCM can maximize the capacity-reach of IM/DD systems. The 100Gbit/s entropy-loaded and clipped multi-rate Nyquist-SCM signal over 100km dispersion-uncompensated link achieves a record capacity-reach of 10Tbit/s$\times$km with BER below $10^{-2}$ and NGMI of 0.932.
$\\$\emph{This work is supported in part by National Key R\&D Program of China (2018YFB1800902); National Natural Science Foundation of China (62005102, U2001601); Natural Science Foundation of Guangdong Province (2019A1515011059); Guangzhou Basic and Applied Basic Research Foundation (202102020996); Fundamental Research Funds for the Central Universities (21619309); Open Fund of IPOC (BUPT) (IPOC2019A001); Hong Kong Scholars Program (XJ2021018).}

\end{document}